\documentclass[twocolumn,amsmath,amssymb,prl,preprintnumbers]{revtex4}
\usepackage{graphicx}% Include figure files
\usepackage{dcolumn}% Align table columns on decimal point
\usepackage{bm}% bold math
\usepackage{amsmath,amssymb}
\usepackage{color}
\usepackage{longtable}
\usepackage{ulem}

\newcommand{\be}{\begin{equation}}
\newcommand{\ee}{\end{equation}}
\newcommand{\bea}{\begin{eqnarray}}
\newcommand{\eea}{\end{eqnarray}}
\newcommand{\bi}{\begin{itemize}}
\newcommand{\ei}{\end{itemize}}

\newcommand\QQbar{q\overline{q}}
\newcommand\mQ{m_{q}}
\newcommand\SdotS{{\mathbf S}_q\cdot{\mathbf S}_{\bar{q}}}

\begin{document}

%-- title ---

\title{
Interquark potential with finite quark mass from lattice QCD}

%-- author list ---

\author{Taichi Kawanai}
\email{kawanai@nt.phys.s.u-tokyo.ac.jp}

\author{Shoichi Sasaki}
\email{ssasaki@phys.s.u-tokyo.ac.jp}
\affiliation{Department of Physics, The University of Tokyo, \\
Hongo 7-3-1, Tokyo 113-0033, Japan}

%-- preprint number ---
\preprint{TKYNT-11-02}

\date{\today}

%-- abstract ---
\begin{abstract}
 We present an investigation of the interquark potential determined
 from the $\QQbar$ Bethe-Salpeter (BS) amplitude for heavy quarkonia 
 in lattice QCD. The $\QQbar$ potential at finite quark mass $\mQ$
 can be calculated from the equal-time and Coulomb gauge BS amplitude 
 through the effective Schr\"odinger equation. The definition of the potential itself requires information 
 about a kinetic mass of the quark.
 We then propose a self-consistent determination of the quark kinetic mass
 on the same footing.
  To verify the proposed method, we perform quenched lattice QCD simulations 
 with a relativistic heavy-quark action at a lattice cutoff of  $1/a\approx 2.1$ GeV in a 
 range of the quark kinetic mass, $1.0 \le \mQ \le 3.6 \;{\rm GeV}$.
 Our numerical results show that the $\QQbar$ potential in the infinitely heavy-quark 
 limit ($\mQ \rightarrow \infty$) is fairly consistent with the conventional one obtained 
 from Wilson loops. The quark-mass dependence of the $\QQbar$ potential
 and the spin-spin potential are also examined.
 \end{abstract}


\maketitle

%--- main text -------------------------------------------------------

%
%intro
%
One of the major successes of lattice QCD is to demonstrate
that a confining potential naturally emerges between 
a static, infinitely heavy quark ($q$) and antiquark ($\bar{q}$)
by considering the Wilson loop~\cite{Bali:2000gf}. This static $\QQbar$ potential can be well 
parametrized by the Coulomb plus linear
potential, which resembles the phenomenology of confining quark interactions,
the so-called Cornell potential~\cite{Eichten:1975ag}, adopted for studying heavy-quark bound systems
such as charmonia and bottomonia~\cite{{Bali:2000gf},{Godfrey:1985xj},{Barnes:2005pb}}. 
However, there is still a gap between the static $\QQbar$ potential 
expressed in terms of Wilson loops
and the phenomenological heavy-quark potential in quark 
models. The former is calculated for infinitely heavy quarks, while the latter is 
exploited at heavy, but finite quark mass.

Recently, the leading and next-to-leading order corrections, which
are classified in powers of the inverse of heavy-quark mass $\mQ$
or powers of the relative quark velocity $v$ within a framework called potential 
nonrelativistic QCD~\cite{Brambilla:2004jw},
to the static $\QQbar$ potential given 
by Wilson loops have been successfully calculated 
in quenched lattice QCD with high accuracy by using a
multilevel algorithm~\cite{Koma:2006si,Koma:2006fw}. Although the $1/\mQ$ (or $v$) expansion
in the Wilson loop formalism could be continued systematically to higher orders, where
the bottom quark can be treated, this methodology is no longer 
applicable at the mass of the charm quark, where 
the motion of the quark and antiquark should not be small enough
to be perturbatively treated~\cite{Bali:2000gf}.

How can we incorporate the effect stemming from the motion of the quark and
antiquark beyond the perturbative treatment?
Ikeda and Iida have recently proposed an interesting idea that the $\QQbar$ 
interquark potential at finite quark mass can be calculated within the Bethe-Salpeter (BS) 
amplitude approach~\cite{Ikeda:2010nj}, which is originally applied for the hadron-hadron interactions including
the first attempt of the nuclear force~\cite{Ishii:2006ec}. 
However, the definition of the interquark potential itself 
{\it requires extra information of the quark mass}.

In this letter, we propose the self-consistent determination of the quark kinetic mass 
within this BS amplitude approach. The quark kinetic mass $m_q$ is defined through the large-distance behavior
in the spin-dependent part of the interquark potential with the help of the measured 
hyperfine splitting of $1S$ states in heavy quarkonia.
Then, we can define a proper interquark potential at finite quark mass, 
which potentially accounts for all orders of $1/\mQ$ corrections.

%%%%%%%%%%%%%%%%%%%%%%%%%%%%%%%%%%%%%%%%%%%%%%%%%%%%%%%%
% formulation
%%%%%%%%%%%%%%%%%%%%%%%%%%%%%%%%%%%%%%%%%%%%%%%%%%%%%%%%
 Let us briefly review the new method utilized here to
 calculate the interquark potential $V_{\QQbar}$ at finite quark mass.
 A gauge-invariant definition of the equal-time $\QQbar$ BS amplitude 
 for quarkonium states is given by
 \begin{equation}
 \phi_{\Gamma}({\bf r})=%\frac{1}{L^3}
 \sum_{{\bf x}}\langle 0| \overline{q}
 ({\bf x})\Gamma {\cal M}({\bf x}, {\bf x}+{\bf r})q({\bf x}+{\bf r})|q\bar{q};J^P\rangle
 \end{equation}
 where ${\bf r}$ is the relative coordinate of two quarks at a certain
 time slice,  $\Gamma$ is 
 chosen as $\gamma_5$ for the pseudoscalar (PS) state
 $(J^{P}=0^-)$,  and $\gamma_i$ for the vector (V) state $(J^{P}=1^{-})$~\cite{{Velikson:1984qw},{Gupta:1993vp}}.
 ${\cal M}$ is a path-ordered product of gauge links.
 In the Coulomb or Landau gauge, the BS amplitude
 can be simply evaluated with ${\cal M}=1$.  
 Hereafter, we consider the Coulomb gauge BS amplitude.
 This amplitude is given by the following four-point correlation function:
 %
 % eq
 %
 \begin{multline}
  G_{\Gamma}({\bf r}, t; t_{\rm src})\\
   =%\sum_{\bf x}\sum_{{\bf x}^{\prime}, {\bf y}^{\prime}}
  \sum_{{\bf x},{\bf x}^{\prime}, {\bf y}^{\prime}}
  \langle \bar{q}({\bf x}, t)\Gamma q({\bf x}+{\bf r}, t) 
   \left(\bar{q}({\bf x}^{\prime}, t_{\rm src})
      \Gamma q({\bf y}^{\prime},t_{\rm src}) \right)^{\dagger}\rangle
  \label{Eq.prop}
 \end{multline}
 where both quark and antiquark at source location ($t_{\rm src}$) 
 are separately projected onto a zero-momentum state by a summation 
 over all spatial coordinates ${\bf x}^{\prime}$ and ${\bf y}^{\prime}$.
 Suppose that $|t-t_{\rm src}|/a\gg 1$
 is satisfied, the four-point correlation function asymptotically behaves as
 \begin{equation}
 G_{\Gamma}({\bf r}, t ; t_{\rm src})\propto
 \phi_{\Gamma}({\bf r}) e^{-M_{\Gamma}(t-t_{\rm src})},
 \end{equation}
 where $M_{\Gamma}$ is the rest mass for the ground state of 
 heavy quarkonia and the ${\bf r}$-dependent amplitude corresponds to
 the Coulomb gauge BS amplitude, namely,  the BS 
 wave function~\cite{{Velikson:1984qw},{Gupta:1993vp}}.    
 After an appropriate projection with respect to discrete rotation, 
 we can get the BS wave function projected in the $A^+_1$ 
 representation, $\phi_{\Gamma}({\bf r})\rightarrow \phi_{\Gamma}(A^{+}_1; r)$, 
 which corresponds to the $S$ wave in continuum theory
 at low energy. Details of the $A_1^{+}$ projection are described in Ref.~\cite{Luscher:1990ux}.
 We simply denote the $A_1^{+}$ projected BS wave function by $\phi_{\Gamma}(r)$
 hereafter.

  The interquark potential $V_{\Gamma}$ can be determined from
  the projected BS wave function $\phi_{\Gamma}(r)$ through
  the stationary Schr\"odinger equation~\footnote{Although the potential defined in this way 
  is generally non-local, it becomes a local potential at low energies
  in a sense of the derivative expansion.
  A detailed discussion can be found in Ref.~\cite{Aoki:2009ji}. }:
 \begin{equation}
  V_{\Gamma}(r)-E_{\Gamma}=\frac{1}{m_q}
   \frac{\nabla^2 \phi_{\Gamma}(r)}
   {\phi_{\Gamma}(r)},
   \label{Eq.Pot}
 \end{equation}
 where {\small$m_q$} is the quark kinetic mass and
 {\small $\nabla^2$} is defined by the discrete Laplacian with
 nearest-neighbor points. The energy eigenvalue $E_{\Gamma}$
 of the stationary Schr\"odinger equation
 is supposed to be $M_{\Gamma}-2m_q$.
 Here we note that this definition of the potential itself requires information
 of the quark mass $m_q$, while the rest mass of the heavy quarkonium state $M_{\Gamma}$
 can be determined by the standard hadron spectroscopy.
 
 The central potential %$V_{\Gamma}$ 
 calculated from $1S$ quarkonium states
 can be decomposed into the spin-independent and spin-dependent parts:
 $V_{\Gamma}(r)=V_{\QQbar}(r)+V_{\rm spin}(r)\SdotS$,
 where $V_{q\bar{q}}$ represents the spin-independent central potential while $V_{\rm spin}$
 corresponds to the spin-spin potential. 
 For the PS and V channels, the spin operator $\SdotS$ 
 can be easily replaced by expectation values $-3/4$ and $1/4$, respectively.
 Therefore,  the potential $V_{q\bar{q}}$ 
 can be evaluated by a linear combination of potentials calculated from the PS and V
 channels as $V_{\QQbar}(r)=\frac{1}{4}[V_{\rm PS}(r)+3V_{\rm V}(r)]$.
 
 As we previously pointed out, the quark kinetic mass $m_q$ is a key ingredient 
 in order to calculate the $\QQbar$ potential defined in Eq.~(\ref{Eq.Pot}) from the BS wave function.
 How can we determine the quark mass?  
 In the initial attempt~\cite{Ikeda:2010nj}, $m_q$
 was approximately evaluated by half of the vector quarkonium mass $M_V/2$.
 However, such an approximate treatment loses a proper quark-mass dependence of the $\QQbar$ potential, 
 which guarantees that the potential defined here is smoothly connected to
 the static $\QQbar$ potential from Wilson loops in the $m_q \rightarrow \infty$ limit.

  %
  % Table
  %  
  \begin{table}%[H]
   \caption{Results of the quark mass $m_q$, 
   the Cornell parameters $A$, $\sigma$,  and the ratio $A/\sigma$   
   in this approach.  
   Their extrapolated values to the
   $m_q \rightarrow \infty$ limit are also compared with 
   the Wilson loop results taken from Ref.~\cite{Koma:2006fw}.
   \label{Tab}
   }
   \begin{ruledtabular}
   \begin{tabular}{|ccccc|} %\hline  \hline 
    $\kappa$ & $am_q$ 
    & $A$ & $a^2\sigma$  &  $A/a^2\sigma$ \\ \hline
    0.11456 &  0.493(18) & 0.663(23)& 0.0477(28) & 13.9(7)\\ 
    0.10190 &  0.833(31) & 0.470(16)& 0.0435(25) & 10.8(6)\\
    0.09495 &  1.006(41) & 0.430(16)& 0.0426(27) & 10.1(6)\\
    0.08333 &  1.288(30) & 0.381(10)& 0.0435(18) & 8.8(4)\\
    0.07490 &  1.484(22) & 0.360(7) & 0.0443(13)  & 8.1(3)  \\
    0.06667 &  1.720(18) & 0.341(6) & 0.0442(11) & 7.7(3)\\
    \hline
   --- & $\infty$  & 0.236(39) & 0.0465(34) & 6.1(1.1)\\ \hline
    \multicolumn{2}{|c}{Wilson loop}  &  0.281(5) & 0.0466(2) &  6.03(11)\\
    %Wilson loop & ---  &  0.281(5) & 0.0466(2) &  6.03(11)\\
  \end{tabular}
  \end{ruledtabular}
\end{table}

  We may alternatively determine the quark mass from the gauge dependent pole mass, 
  which can be measured by the quark two-point function {\it in the Landau gauge}. 
  We instead propose a novel method which is applicable even
  {\it in the Coulomb gauge} as follows.
  We first consider the spin-dependent potential, which is given by
   \begin{equation}
    V_{\rm spin}(r)-\Delta E_{\rm hyp}=\frac{1}{m_q}
     \left(\frac{\nabla^2 \phi_{\rm V}(r)}{\phi_{\rm V}(r)}
      -\frac{\nabla^2 \phi_{\rm PS}(r)}{\phi_{\rm PS}(r)}\right),
     \label{Eq_Pot_hyp}
   \end{equation}
   where
   $\Delta E_{\rm hyp}$ denotes a difference between
   energy eigenvalues of the PS and V channels. 
   Indeed, the value of $\Delta E_{\rm hyp}$ is nothing but
   hyperfine mass splitting $M_{\rm V}-M_{\rm PS}$.

   %Suppose that $\lim_{r\to \infty}V_{\rm spin}(r)=0$,
   According to the $1/\mQ$ expansion approach, 
   we may expect to have
   the condition of $\lim_{r\to \infty}V_{\rm spin}(r)=0$,
   which implies that there is no long-range correlation and
   no irrelevant constant term in the spin-dependent potential~\cite{
   {Brambilla:2004jw},{Koma:2006fw}}.
   We thus can estimate the quark kinetic mass $m_q$ through 
   the following formula:
   %
   % quark kinetic mass
   % 
   \begin{equation}
    m_q = \lim_{r\to \infty}\frac{1}{\Delta E_{\rm hyp}}
     \left(\frac{\nabla^2 \phi_{\rm PS}(r)}{\phi_{\rm PS}(r)}
      -\frac{\nabla^2 \phi_{\rm V}(r)}{\phi_{\rm V}(r)}\right),
   \label{Eq:KinQmass}
   \end{equation}
   where $\Delta E_{\rm hyp}=M_{\rm V}-M_{\rm PS}$ is measured by the standard 
   hadron spectroscopy.
   As a result, one can self-consistently determine both the spin-independent and spin-dependent $\QQbar$ potentials,
   and also the quark kinetic mass within a single set of four-point correlation functions.

 %%%%%%%%%%%%%%%%%%%%%%%%%%%%%%%%%%%%%%%%%%%%%%%%%%%%%%
 % numerical results
 %%%%%%%%%%%%%%%%%%%%%%%%%%%%%%%%%%%%%%%%%%%%%%%%%%%%%%
 To verify our new proposal, we have performed quenched lattice QCD simulations on a 
 lattice $L^3\times T=32^3\times 48$ with the standard single-plaquette gauge action at
 $\beta=6/g^2=6.0$, which corresponds to a lattice cutoff of 
 $a^{-1}\approx 2.1$ GeV ($a \approx 0.093$fm).
 %according to the Sommer scale.
 The spatial lattice size corresponds to $La \approx 3\;{\rm fm}$.
 We fix the lattice to Coulomb gauge.
 The heavy-quark propagators are computed using the relativistic heavy-quark (RHQ) action 
 with relevant one-loop coefficients of the RHQ~\cite{{Aoki:2001ra},{Kayaba:2006cg}}.
 The RHQ action utilized here is a variant of the Fermilab approach~\cite{ElKhadra:1996mp}
 and can remove large discretization errors introduced by large quark mass.

 To examine the infinitely heavy-quark limit, 
 we adopt the six values of the hopping parameter $\kappa$,
 %$\kappa_q=\{0.11456,\ 0.10190,\ 0.09495,\ 0.08333,\ 0.07490,\ 0.06333\}$, 
 which cover the range of the spin-averaged mass of $1S$ quarkonium states 
 $M_{\rm ave}=\frac{1}{4} (M_{\rm PS} + 3 M_{\rm V})=1.97$-5.86 GeV.
  We calculate quark propagators with a wall source
  which is located at $t_{\text{src}}/a=4$.
  Dirichlet boundary conditions are imposed for time direction.
  Our results are analyzed on 150 configurations for every hopping parameters.
  In this letter, we use only the on-axis data of the BS wave function since the off-axis 
  data may suffer more from the rotational symmetry breaking effect.

 %
 %Results1
 %
 %
 %
  \begin{figure}
   \centering
   \includegraphics[width=.49\textwidth]{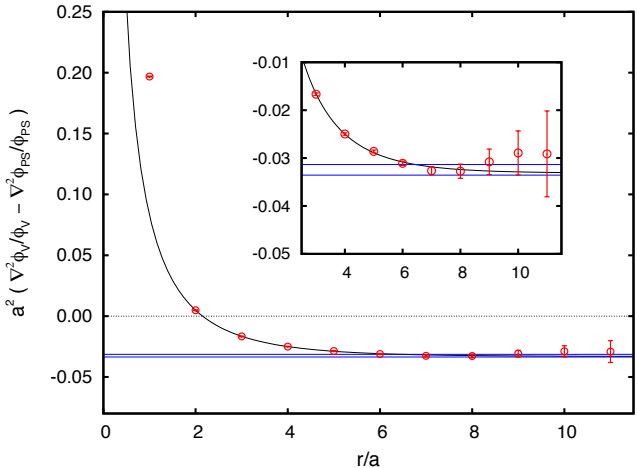}
   \caption{A typical result of 
   $\nabla^2\phi_{\rm V}/\phi_{\rm V}-\nabla^2\phi_{\rm PS}/\phi_{\rm PS}$
   as a function of spacial distance $r$,
   which is calculated at a quark mass close to the charm quark mass. 
   A nonzero constant shift observed at large distances
   corresponds to the value of $-m_q\Delta E_{\rm hyp}$ (solid line), which is
   determined by a weighted average of data points 
   at $r/a=7$~-~11 in this case.
   A solid curve is a fit to the Yukawa plus constant form.
   \label{Pot_hyp}}
  \end{figure}

  First, in Fig.~\ref{Pot_hyp}, we plot a difference of ratios of 
  $\nabla^2\phi_{\rm V}/\phi_{\rm V}$ and $\nabla^2\phi_{\rm PS}/\phi_{\rm PS}$
  as a function of spatial distance $r$ at $\kappa=0.10190$, which is close to the charm quark mass~\cite{Kawanai:2010ev},
  as a typical example. The ratios of $\nabla^2\phi_{\Gamma}/\phi_{\Gamma}$ are 
  evaluated by a weighted average of data points in the
  range of $(t-t_{\rm src})/a=$~21~-~23. 
  At a glance, the value of $\nabla^2\phi_{\rm V}/\phi_{\rm V}-\nabla^2\phi_{\rm PS}/\phi_{\rm PS}$   
  certainly reaches a nonzero constant value at large distances, which turns out to be 
  the value of $-m_q\Delta E_{\rm hyp}$.   
  %This observation is 
  %enough to validate our assumption of $\lim_{r\to \infty}V_{\rm spin}(r)=0$.
  The values of $m_q\Delta E_{\rm hyp}$ 
  are evaluated by a weighted average of 
  data points in the range 
  where $V_{\rm spin}(r)$ should vanish.
  %must vanish.
  We then obtain the quark kinetic masses from 
  the long-distance asymptotic values of 
  $\nabla^2\phi_{\rm V}/\phi_{\rm V}-\nabla^2\phi_{\rm PS}/\phi_{\rm PS}$ 
  divided by the measured hyperfine splitting $\Delta E_{\rm hyp}$. 
  
  Apart from the vertical scale and offset, Fig.~\ref{Pot_hyp} exhibits
  the $r$ dependence of the spin-spin potential $V_{\rm spin}(r)$. 
  The potential is quickly dumped at large distances as mentioned previously.
  Although the spin-spin interaction based on one-gluon exchange
  like the Fermi-Breit interaction of QED is described by a short-range
  $\delta$-function potential, it is evident that the obtained spin-spin potential
  is a repulsive potential with some finite range. Indeed, the long-range screening
  observed here is simply accommodated by exponential-type fitting functions 
  $\exp(-r^m)/r^n$.  %where $n$ and $m$ are not restricted to be integers. 
  If we adopt the Yukawa form ($m=n=1$) with a constant shift
  $c e^{-\alpha r}/r + d$ to fit our data of $V_{\rm spin}(r)$,
  we can get a reasonable fit 
  over the range of $r/a$ from $2$ to $11$. 
  The fitting result is displayed as a solid curve in Fig.\ref{Pot_hyp}.
  In addition, the parameter $\alpha$ is very sensitive 
  to the quark mass. It increases  with the quark mass  
  in the measured mass region of $m_q=1.0$~-~3.6 GeV.
  We plan to present more detail on the spin-spin 
  potential in a separate publication.

  %
  %Results2
  %
  %
  \begin{figure}
   \centering
   \includegraphics[width=.49\textwidth]{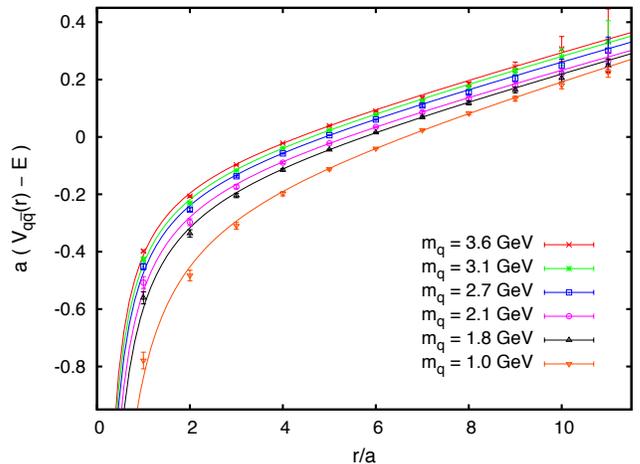}
   \caption{The interquark potential calculated from the $\QQbar$ BS amplitude
   at finite quark masses covering the range from 1.0 to 3.6~GeV. 
   For clarity of the figure, the constant energy shift $E$ is not subtracted. 
   Each curve represents the fit result with the Cornell parametrization defined
   in Eq.~(\ref{Eq_QQPot}).}
   \label{Pot}
   \end{figure}

  Using the quark kinetic mass determined here, we can properly calculate
  the spin-independent interquark potential $V_{\QQbar}(r)$ from the BS wave functions
  without any ambiguity.
  Figure~\ref{Pot} displays results of our potential 
  $V_{\QQbar}$ obtained at several quark masses. For clarity of the figure, 
  the constant energy shift $E$, which corresponds to the value of $M_{\rm av}-2m_q<0$, 
  is not subtracted in Fig.~\ref{Pot}. 
  The resulting $\QQbar$ potentials at finite quark masses exhibit
  the linearly rising potential at large distances and the Coulomb-like potential at short distances
  as originally reported in Ref~\cite{Ikeda:2010nj}.

  We simply adopt the Cornell parametrization for fitting our data:
  %
  %Eq
  %
  \begin{equation}
   V_{q\bar{q}}(r)=-\frac{A}{r}+\sigma r + V_0
    \label{Eq_QQPot}
  \end{equation}
  with the Coulombic coefficient~$A$, the string tension $\sigma$, 
  and a constant $V_0$.
  It is found that the Cornell potential describes well the $\QQbar$
  potential even at a lighter quark mass than the charm quark mass.
  Although such a light mass region is beyond the radius of convergence for the systematic 
  $1/\mQ$ expansion, the finite $m_q$ corrections could be nonperturbatively 
  encoded into the Cornell parameters in this approach. 
  All fits are performed over the range $1\le r/a \le 11$.
  Fit results for $A$ and $\sigma$ are summarized in Table~\ref{Tab}.

\begin{figure}
\centering
\includegraphics[width=.49\textwidth]{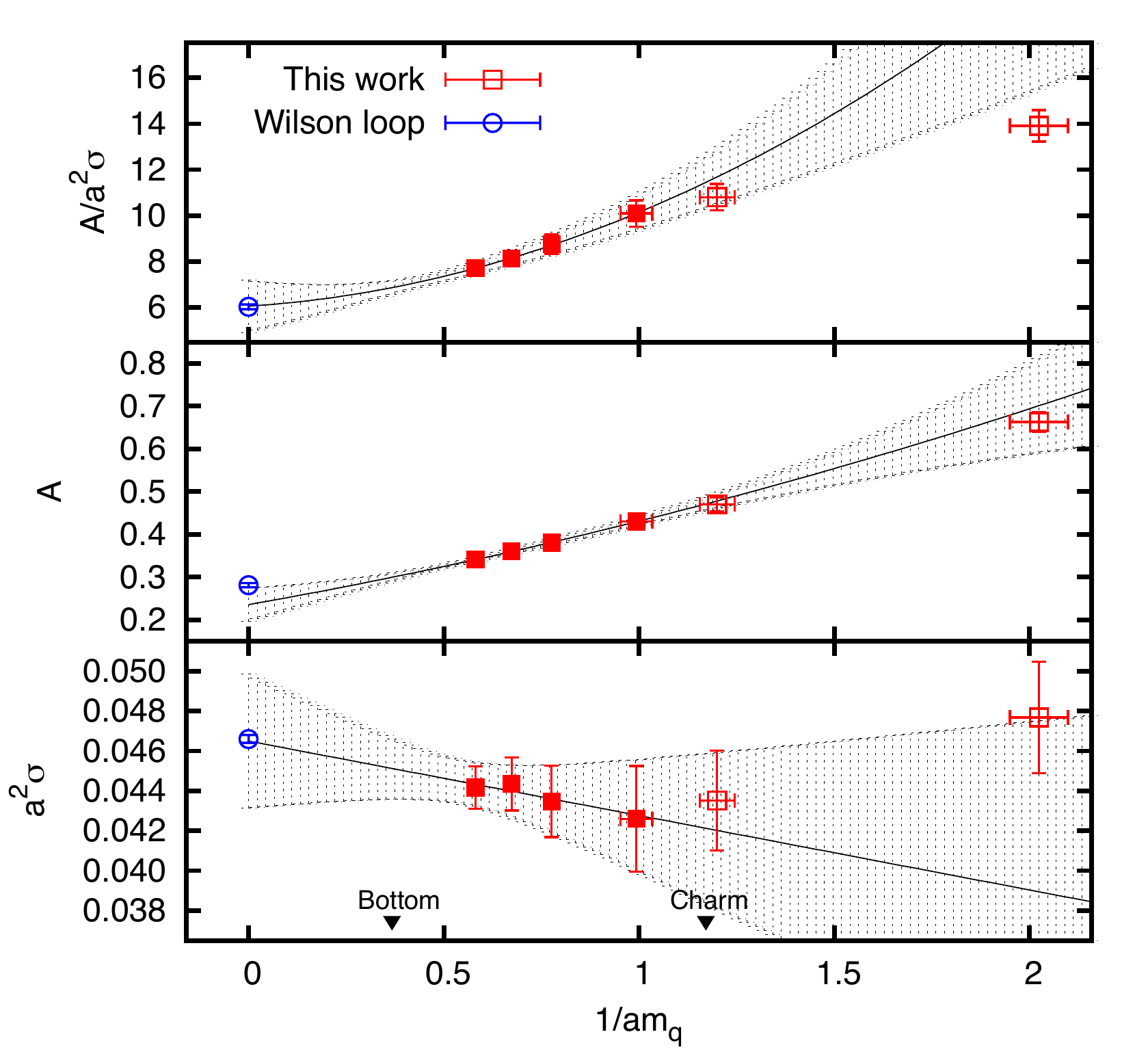}
\caption{
The quark-mass dependence of $A/\sigma$, $A$, and $\sigma$
as functions of $1/\mQ$. We perform the extrapolation towards
the $m_q \rightarrow \infty$ limit (solid curves) of $A/\sigma$, $A$,  and $\sigma$
with a simple polynomial function in $1/\mQ$.
For $\sigma$, a linear fit with respect to $1/\mQ$ is enough
to describe the data with reasonable $\chi^2/{\rm d.o.f}$, while
quadratic fits are used for $A/\sigma$ and $A$.
The results given by Wilson loops  
are also included as open circles. .
\label{para}}
\end{figure}

  In Fig.~\ref{para}, we show the quark-mass dependence of $A/\sigma$, $A$ and $\sigma$
  as functions of $1/\mQ$. First, regardless of the definition of $\mQ$, 
  the ratio of $A/\sigma$ in the top figure indicates that the $\QQbar$ potential calculated from the BS wave
  function smoothly approaches the potential obtained from Wilson loops in the infinitely heavy-quark limit.
  Indeed, the $\mQ \rightarrow \infty$ extrapolation by a quadratic fit (solid curve) with respect to $1/\mQ$ 
  to four heaviest points (filled squares), where $1/\mQ$ in lattice units is less than unity, 
  is consistent with the value obtained from Wilson loops. 
   
  If we pay attention to the quark-mass dependence of each of the Cornell parameters separately, 
  we observe that,  although the Coulombic parameter $A$ depends on
  the quark mass significantly, there is no appreciable dependence of the quark mass
  on the string tension $\sigma$. These observations agree with several features of the $1/\mQ$ corrections
  to the static $\QQbar$ potential found in Refs.~\cite{Koma:2006si,Koma:2006fw}.  
  The long-range part of the potential characterized by 
  the string tension receives no correction up to ${\cal O}(1/m_q^2)$ in the Wilson loop formalism, 
  while the $1/\mQ$ correction on the Coulomb term starts at ${\cal O}(1/m_q)$.
  Here we recall that our way of determining the interquark potential with the proper quark mass defined 
  in Eq.~(\ref{Eq:KinQmass}) is responsible for desirable quark-mass dependence observed here. 
  
  We finally evaluate the values of $A$ and $\sigma$ in the infinitely heavy-quark limit
  by quadratic and linear fits with respect to $1/\mQ$, respectively. 
  Extrapolated values at $m_q \rightarrow \infty$ are again consistent with those of the 
  Wilson loop result. 
  The extrapolation curve and line are also displayed as a solid curve and line in Fig.~\ref{para}.

%

  %
  %summary
  %
  In this letter, we have proposed the new method to determine 
  the interquark potential $V_{\QQbar}(r)$ at finite quark mass for 
  heavy quarkonia from lattice QCD.
  The self-consistent determination of the quark kinetic mass $\mQ$ plays an essential
  role on how to define the interquark potential from 
  the $\QQbar$ BS amplitude. Using quenched lattice QCD, we have demonstrated
  that the spin-independent central potential defined in this method
  smoothly approaches the static $\QQbar$ potential given by Wilson loops 
  in the infinitely heavy-quark limit.
  We also found that our defined potential, which is well described by the Cornell parametrization 
  even at finite quark mass, exhibits the desired quark-mass dependence of the Cornell parameters, 
  which agrees well with what is expected from results obtained in the systematic $1/\mQ$ expansion approach. 
  This suggests that our interquark potential properly accounts for all orders of $1/\mQ$ corrections and
  its nature is validated at least up to the charm quark sector.
  We also stress that this new method allows us to calculate
  the spin-dependent potential with high accuracy. 
  We will next apply to dynamical calculation of 
  the interquark potential at the charm quark mass, of which precise knowledge is 
  important especially for physics of higher charmonia~\cite{Barnes:2005pb}. Such planning is now underway.

 %
 %acknowledgement
 %
  We would like to thank T. Hatsuda for helpful suggestions, H. Iida and Y. Ikeda for fruitful
  discussions.
    T.K. is supported by Grant-in-Aid for the Japan Society for Promotion
  of Science (JSPS) (No.~22-7653).
  S.S. is  supported by the JSPS Grant-in-Aids for Scientific Research (C)
  (No.~19540265) and Scientific Research on Innovative Areas
  (No.~21105504).


\begin{references}

  %\cite{Bali:2000gf}
  \bibitem{Bali:2000gf}
  For a review, see G.~S.~Bali,
  %``QCD forces and heavy quark bound states,''
  Phys.\ Rept.\  {\bf 343}, 1 (2001).
  %[arXiv:hep-ph/0001312].
  %%CITATION = PRPLC,343,1;%%

  %\cite{Eichten:1975ag}
  \bibitem{Eichten:1975ag}
  E.~Eichten {\it et.al.}, %K.~Gottfried, T.~Kinoshita, K.~D.~Lane and T.~M.~Yan,
  %``The Interplay Of Confinement And Decay In The Spectrum Of Charmonium,''
  Phys.\ Rev.\ Lett.\  {\bf 36}, 500 (1976).
  %%CITATION = PRLTA,36,500;%%
  
  %\cite{Godfrey:1985xj}
  \bibitem{Godfrey:1985xj}
  S.~Godfrey and N.~Isgur,
  %``Mesons In A Relativized Quark Model With Chromodynamics,''
  Phys.\ Rev.\  D {\bf 32}, 189 (1985).
  %%CITATION = PHRVA,D32,189;%%

  %\cite{Barnes:2005pb}
  \bibitem{Barnes:2005pb}
  T.~Barnes, S.~Godfrey and E.~S.~Swanson,
  %``Higher Charmonia,''
  Phys.\ Rev.\  D {\bf 72}, 054026 (2005).
  %[arXiv:hep-ph/0505002].
  %%CITATION = PHRVA,D72,054026;%%
  
%%%% pNRQCD 

%\cite{Brambilla:2004jw}
\bibitem{Brambilla:2004jw}
  %For a review, see
  N.~Brambilla {\it et al.}, %A.~Pineda, J.~Soto and A.~Vairo,
  %``Effective field theories for heavy quarkonium,''
  Rev.\ Mod.\ Phys.\  {\bf 77}, 1423 (2005).
  %[arXiv:hep-ph/0410047].
  %%CITATION = RMPHA,77,1423;%%
    
  %\cite{Koma:2006si}
  \bibitem{Koma:2006si}
  Y.~Koma, M.~Koma and H.~Wittig,
  %``Nonperturbative determination of the QCD potential at O(1/m),''
  Phys.\ Rev.\ Lett.\  {\bf 97}, 122003 (2006).
  %[arXiv:hep-lat/0607009].
  %%CITATION = PRLTA,97,122003;%%

  %\cite{Koma:2006fw}
  \bibitem{Koma:2006fw}
  Y.~Koma and M.~Koma,
  %``Spin-dependent potentials from lattice QCD,''
  Nucl.\ Phys.\  B {\bf 769}, 79 (2007).
  %[arXiv:hep-lat/0609078].
  %%CITATION = NUPHA,B769,79;%%


  %\cite{Ikeda:2010nj}
  \bibitem{Ikeda:2010nj}
  Y.~Ikeda and H.~Iida,
  %``The anti-quark--quark potential from Bethe-Salpeter amplitudes on
  %lattice,''
  arXiv:1011.2866 %[hep-lat],
  %%CITATION = ARXIV:1011.2866;%%
  %\cite{Ikeda:2011bs}
  %\bibitem{Ikeda:2011bs}
  %Y.~Ikeda and H.~Iida,
  %``Quark-anti-quark potentials from Nambu-Bethe-Salpeter amplitudes on
  %lattice,''
  and 1102.2097. %[hep-lat].
  %%CITATION = ARXIV:1102.2097;%%

  %\cite{Ishii:2006ec}
  \bibitem{Ishii:2006ec}
  N.~Ishii, S.~Aoki and T.~Hatsuda,
  %``The nuclear force from lattice QCD,''
  Phys.\ Rev.\ Lett.\  {\bf 99}, 022001 (2007).
  %[arXiv:nucl-th/0611096].
  %%CITATION = PRLTA,99,022001;%%
  
  %\cite{Aoki:2009ji}
  \bibitem{Aoki:2009ji}
  S.~Aoki, T.~Hatsuda and N.~Ishii,
  %``Theoretical Foundation of the Nuclear Force in QCD and its applications to
  %Central and Tensor Forces in Quenched Lattice QCD Simulations,''
  Prog.\ Theor.\ Phys.\  {\bf 123} (2010) 89.
  %[arXiv:0909.5585 [hep-lat]].
  %%CITATION = PTPKA,123,89;%%

%%% qqbar BS amplitude %%%%
  %\cite{Velikson:1984qw}
  \bibitem{Velikson:1984qw}
  B.~Velikson and D.~Weingarten,
  %``Hadron Wave Functions In Lattice QCD,''
  Nucl.\ Phys.\  B {\bf 249}, 433 (1985).
  %%CITATION = NUPHA,B249,433;%%
  %\cite{Gupta:1993vp}

  \bibitem{Gupta:1993vp}
  R.~Gupta {\it et al.}, %D.~Daniel and J.~Grandy,
  %``Bethe-Salpeter amplitudes and density correlations for mesons with Wilson
  %fermions,''
  Phys.\ Rev.\  D {\bf 48}, 3330 (1993).
  %[arXiv:hep-lat/9304009].
  %%CITATION = PHRVA,D48,3330;%%
  
  %\cite{Luscher:1990ux}
  \bibitem{Luscher:1990ux}
  M.~L\"uscher,
  %``Two Particle States On A Torus And Their Relation To The Scattering
  %Matrix,''
  Nucl.\ Phys.\ B {\bf 354}, 531 (1991).
  %%CITATION = NUPHA,B354,531;%%    

  %\cite{Aoki:2001ra}
  \bibitem{Aoki:2001ra}
  S.~Aoki, Y.~Kuramashi and S.~I.~Tominaga,
  %``Relativistic heavy quarks on the lattice,''
  Prog.\ Theor.\ Phys.\  {\bf 109}, 383 (2003).
  %[arXiv:hep-lat/0107009].
  %%CITATION = PTPKA,109,383;%%

  %\cite{Kayaba:2006cg}
  \bibitem{Kayaba:2006cg}
  Y.~Kayaba {\it et al.}  [CP-PACS Collaboration],
  %``First Nonperturbative Test of a Relativistic Heavy Quark Action in Quenched
  %Lattice QCD,''
  JHEP {\bf 0702}, 019 (2007).
  %[arXiv:hep-lat/0611033].
  %%CITATION = JHEPA,0702,019;%%

  %\cite{ElKhadra:1996mp}
  \bibitem{ElKhadra:1996mp}
  A.~X.~El-Khadra {\it et al.}, %, A.~S.~Kronfeld and P.~B.~Mackenzie,
  %``Massive Fermions in Lattice Gauge Theory,''
  Phys.\ Rev.\  D {\bf 55}, 3933 (1997).
  %[arXiv:hep-lat/9604004].
  %%CITATION = PHRVA,D55,3933;%%
  
  %\cite{Kawanai:2010ev}
  \bibitem{Kawanai:2010ev}
  T.~Kawanai and S.~Sasaki,
  %``Charmonium-nucleon potential from lattice QCD,''
  Phys.\ Rev.\  D {\bf 82}, 091501 (2010).
  %[arXiv:1009.3332 [hep-lat]].
  %%CITATION = PHRVA,D82,091501;%%
  
 \end{references}
\end{document}